\documentclass{article}

\usepackage{color}
\usepackage{amsmath, amssymb}

\newcommand{\ZM}{\mathbb{Z}}

\newcommand{\CM}{\mathbb{C}}
\newtheorem{theorem}{Theorem}
 
\newtheorem{prop}{Proposition} 
\newtheorem{cor}{Corollary}

\newcommand{\bec}[1]{\mbox{\boldmath $#1$}}

\begin{document}

\title{{\bf On the relation between \\
quantum walks and absolute zeta functions}
\vspace{15mm}}

\author{Norio KONNO \\ \\
Department of Mathematical Sciences \\
College of Science and Engineering \\
Ritsumeikan University \\
1-1-1 Noji-higashi, Kusatsu, 525-8577, JAPAN \\
e-mail: n-konno@fc.ritsumei.ac.jp \\
}

\date{\empty }

\maketitle

\vspace{50mm}


\vspace{20mm}


\begin{small}
{\bf Abbr. title: Quantum walks and absolute zeta functions} 
\end{small}









\clearpage

\begin{abstract}
The quantum walk is a quantum counterpart of the classical random walk. On the other hand, the absolute zeta function can be considered as a zeta function over $\mathbb{F}_1$. This paper presents a connection between the quantum walk and the absolute zeta function. First we deal with a zeta function determined by a time evolution matrix of the Grover walk on a graph. The Grover walk is a typical model of the quantum walk. Then we prove that the zeta function given by the quantum walk is an absolute automorphic form of weight depending on the number of edges of the graph. Furthermore we consider an absolute zeta function for the zeta function based on a quantum walk. As an example, we compute an absolute zeta function for the cycle graph and show that it is expressed as the multiple gamma function of order 2.
\end{abstract}

\vspace{10mm}

\begin{small}
\end{small}

\vspace{10mm}

\section{Introduction \label{sec01}}
The quantum walk (QW) is considered to be the corresponding model for the random walk (RW) in quantum systems. QWs play important roles in various fields such as mathematics, quantum physics, and quantum information processing. Concerning QW, see \cite{GZ, Konno2008, ManouchehriWang, Portugal, Venegas}, and as for RW, see \cite{Norris, Spitzer}, for instance. On the other hand, the absolute zeta function is a zeta function over $\mathbb{F}_1$, where $\mathbb{F}_1$ can be viewed as a kind of limit of $\mathbb{F}_p$ as $p \to 1$. Here $\mathbb{F}_p = \mathbb{Z}/p \mathbb{Z}$ stands for the field of $p$ elements for a prime number $p$. This paper presents a connection between QWs and absolute zeta functions. Concerning absolute zeta function, see \cite{CC, KF1, Kurokawa3, Kurokawa, KO, KT3, KT4, Soule}, 

In this paper, first we deal with a zeta function $\zeta_{{\bf U}_{G}} (u)$ determined by ${\bf U}_{G}$ which is a time evolution matrix of the Grover walk on $G$, where $G$ is a simple connected graph with $n$ vertices and $m$ edges. Here the Grover walk given by the Grover matrix is a well-studied model in QW research. Then we prove that $\zeta_{{\bf U}_{G}} (u)$ is an absolute automorphic form of weight $- 2m$. Afterwards we consider an absolute zeta function $\zeta_{\zeta_{{\bf U}_{G}}} (s)$ for our zeta function $\zeta_{{\bf U}_{G}} (u)$. As an example, we calculate $\zeta_{\zeta_{{\bf U}_{G}}} (s)$ for the cycle graph $G =C_n$ with $n$ vertices and $n$ edges, and show that it is expressed as the multiple gamma function of order 2 via the multiple Hurwitz zeta function of order 2. Finally, we obtained the functional equation for $\zeta_{\zeta_{{\bf U}_{C_n}}} (s)$ with the multiple sine function of order 2. The present manuscript is the first step of the study on a relationship between the QW and the absolute zeta function.

The rest of this paper is organized as follows. Section \ref{sec02} briefly describes the absolute zeta function and its some related topics. In Section \ref{sec03}, we deal with the definition of the QW and the Konno-Sato theorem. Section \ref{sec04} explains a connection between the QW and the absolute zeta function. In Section \ref{sec05}, we give an example for the cycle graph. Finally, Section \ref{sec06} is devoted to conclusion.

\section{Absolute Zeta Function \label{sec02}}
First we introduce the following notation: $\mathbb{Z}$ is the set of integers, $\mathbb{Z}_{>} = \{1,2,3, \ldots \}$,  $\mathbb{R}$ is the set of real numbers, and $\mathbb{C}$ is the set of complex numbers. 

In this section, we briefly review the framework on the absolute zeta functions, which can be considered as zeta function over $\mathbb{F}_1$, and absolute automorphic forms (see \cite{Kurokawa3, Kurokawa, KO, KT3, KT4} and references therein, for example). 

Let $f(x)$ be a function $f : \mathbb{R} \to \mathbb{C} \cup \{ \infty \}$. We say that $f$ is an {\em absolute automorphic form} of weight $D$ if $f$ satisfies
\begin{align*}
f \left( \frac{1}{x} \right) = C x^{-D} f(x)
\end{align*}
with $C \in \{ -1, 1 \}$ and $D \in \mathbb{Z}$. The {\em absolute Hurwitz zeta function} $Z_f (w,s)$ is defined by
\begin{align*}
Z_f (w,s) = \frac{1}{\Gamma (w)} \int_{1}^{\infty} f(x) \ x^{-s-1} \left( \log x \right)^{w-1} dx,
\end{align*}
where $\Gamma (x)$ is the gamma function (see \cite{Andrews1999}, for instance). Then taking $u=e^t$, we see that $Z_f (w,s)$ can be rewritten as the Mellin transform: 
\begin{align}
Z_f (w,s) = \frac{1}{\Gamma (w)} \int_{0}^{\infty} f(e^t) \ e^{-st} \ t^{w-1} dt.
\label{kirishima01}
\end{align}
Moreover, we introduce the {\em absolute zeta function} $\zeta_f (s)$ as follows:
\begin{align*}
\zeta_f (s) = \exp \left( \frac{\partial}{\partial w} Z_f (w,s) \Big|_{w=0} \right).
\end{align*}
\par
Now we give an example of $f(x)$ which will be used in Section \ref{sec05}:
\begin{align}
f(x) = \frac{1}{(x^N -1)^2}
\label{kirishima02}
\end{align}
with $N \in \mathbb{Z}_{>}$. Then we find 
\begin{align*}
f \left( \frac{1}{x} \right) = x^{2N} f(x),
\end{align*}
so $f$ (given by Eq. \eqref{kirishima02}) is an absolute automorphic form of weight $-2N$. From Eq. \eqref{kirishima01}, we compute
\begin{align*}
Z_f (w,s) 
&= \frac{1}{\Gamma (w)} \int_{0}^{\infty} \frac{e^{-st}}{(e^{Nt} -1)^2} \ t^{w-1} dt
\\
&= \frac{1}{\Gamma (w)} \int_{0}^{\infty} e^{-st} \ \left( \sum_{n_1=1}^{\infty} e^{-n_1 N t} \right) \left( \sum_{n_2=1}^{\infty} e^{-n_2 N t} \right) \ t^{w-1}dt
\\
&= \sum_{n_1=0}^{\infty} \sum_{n_2=0}^{\infty} \int_{0}^{\infty} \frac{1}{\Gamma (w)} e^{- \{s + 2N + (n_1+n_2)N \}t} \ t^{w-1} dt,
\\
&= \sum_{n_1=0}^{\infty} \sum_{n_2=0}^{\infty} \{s + 2N + (n_1+n_2)N \}^{-w}.
\end{align*}
Thus we have 
\begin{align}
Z_f (w,s) = \sum_{n_1=0}^{\infty} \sum_{n_2=0}^{\infty} \{s + 2N + (n_1+n_2)N \}^{-w}.
\label{kaibutsu01}
\end{align}
Here we introduce the {\em multiple Hurwitz zeta function of order $r$}, $\zeta_r (s, x, (\omega_1, \ldots, \omega_r))$, and the {\em multiple gamma function of order $r$}, $\Gamma_r (x, (\omega_1, \ldots, \omega_r))$, in the following (see \cite{Kurokawa3, Kurokawa, KT3}): 
\begin{align}
\zeta_r (s, x, (\omega_1, \ldots, \omega_r))
&= \sum_{n_1=0}^{\infty} \cdots \sum_{n_r=0}^{\infty} \left( n_1 \omega_1 + \cdots + n_r \omega_r + x \right)^{-s}, 
\label{kirishima03}
\\
\Gamma_r (x, (\omega_1, \ldots, \omega_r)) 
&= \exp \left( \frac{\partial}{\partial s} \zeta_r (s, x, (\omega_1, \ldots, \omega_r)) \Big|_{s=0} \right).
\label{kirishima04}
\end{align}
Therefore, combining Eq. \eqref{kaibutsu01} with Eqs. \eqref{kirishima03} and \eqref{kirishima04} yields
\begin{align}
Z_f (w,s) &= \zeta_2 (w, s+2N, (N,N)), 
\label{kirishima05}
\\
\zeta_f (s) &= \Gamma_2 (s+2N, (N,N)). 
\label{kirishima06}
\end{align}
So we see that $Z_f (w,s)$ and $\zeta_f (s)$ can be obtained by a multiple Hurwitz zeta function of order $2$ and a multiple gamma function of order $2$, respectively.

\section{QW and the Konno-Sato Theorem \label{sec03}}
First we deal with the definition of a QW. Afterwards we explain the Konno-Sato Theorem presented in \cite{KonnoSato}. This theorem treats a relation for eigenvalues between QWs and RWs. More precisely, the QW is the Grover walk  determined by the Grover matrix with flip-flop shift type (called F-type), and the RW is a simple symmetric RW whose walker jumps to each of its nearest neighbors with equal probability on a graph. We assume that all the graphs are simple (i.e., without multiple edges and loops) and connected.

Let $G=(V(G),E(G))$ be a simple connected graph with the set $V(G)$ of vertices and the set $E(G)$ of unoriented edges $uv$ joining two vertices $u$ and $v$. Moreover, let $n=|V(G)|$ and $m=|E(G)|$ be the number of vertices and edges of $G$, respectively. For $uv \in E(G)$, an arc $(u,v)$ is the oriented edge from $u$ to $v$. 
Put $D(G)= \{ (u,v),(v,u) \mid uv \in E(G) \}$. For $e=(u,v) \in D(G)$, set $u=o(e)$ and $v=t(e)$. Furthermore, let $e^{-1}=(v,u)$ be the {\em inverse} of $e=(u,v)$. For $v \in V(G)$, the {\em degree} $\deg {}_G \ v = \deg v = d_v $ of $v$ is the number of vertices adjacent to $v$ in $G$. If $ \deg {}_G \ v=k$ (constant) for each $v \in V(G)$, then $G$ is called {\em $k$-regular}. A {\em path $P$ of length $n$} in $G$ is a sequence $P=(e_1, \ldots ,e_n )$ of $n$ arcs such that $e_i \in D(G)$, $t( e_i )=o( e_{i+1} ) \ (1 \leq i \leq n-1)$. If $e_i =( v_{i-1} , v_i )$ for $i=1, \cdots , n$, then we write $P=(v_0, v_1, \ldots ,v_{n-1}, v_n )$. Put $ \mid P \mid =n$, $o(P)=o( e_1 )$ and $t(P)=t( e_n )$. Also, $P$ is called an {\em $(o(P),t(P))$-path}. We say that a path $P=( e_1 , \ldots , e_n )$ has a {\em backtracking} if $ e^{-1}_{i+1} =e_i $ for some $i \ (1 \leq i \leq n-1)$. A $(v, w)$-path is called a {\em $v$-cycle} (or {\em $v$-closed path}) if $v=w$. Let $B^r$ be the cycle obtained by going $r$ times around a cycle $B$. Such a cycle is called a {\em multiple} of $B$. A cycle $C$ is {\em reduced} if both $C$ and $C^2 $ have no backtracking. The {\em Ihara zeta function} of a graph $G$ is a function of a complex variable $u$ with $|u|$ sufficiently small, defined by 
\begin{align*}
{\bf Z} (G, u)= \exp \left( \sum^{\infty}_{r=1} \frac{N_r}{r} u^r \right), 
\end{align*}
where $N_r$ is the number of reduced cycles of length $r$ in $G$. Let $G$ be a simple connected graph with $n$ vertices $v_1, \ldots ,v_n $. The {\em adjacency matrix} ${\bf A}_n = [a_{ij} ]$ is the $n \times n$ matrix such that $a_{ij} =1$ if $v_i$ and $v_j$ are adjacent, and $a_{ij} =0$ otherwise. The following result is due to Ihara \cite{Ihara} and Bass \cite{Bass}.
\begin{theorem}
Let $G$ be a simple connected graph with $V(G)= \{ v_1 , \ldots , v_n \}$ and $m$ edges. Then we have
\begin{align*}
{\bf Z} (G,u )^{-1} =(1- u^2 )^{\gamma-1} 
\det \left( {\bf I}_n -u {\bf A}_n + u^2 ( {\bf D}_n - {\bf I}_n ) \right). 
\end{align*}
Here $\gamma$ is the Betti number of $G$ {\rm (}i.e., $\gamma = m - n +1${\rm )}, ${\bf I}_n$ is the $n \times n$ identity matrix, and ${\bf D}_n = [d_{ij}]$ is the $n \times n$ diagonal matrix with $d_{ii} = \deg v_i$ and $d_{ij} =0 \ (i \neq j)$. 
\end{theorem}

Let $G$ be a simple connected graph with $V(G)= \{ v_1 , \ldots , v_n \}$ and $m$ edges. Then the $2m \times 2m$ {\em Grover matrix} ${\bf U}_{2m} = [ U_{ef} ]_{e,f \in D(G)} $ of $G$ is defined by 
\begin{align}
U_{ef} =\left\{
\begin{array}{ll}
2/d_{t(f)} (=2/d_{o(e)} ) & \mbox{if $t(f)=o(e)$ and $f \neq e^{-1} $, } \\
2/d_{t(f)} -1 & \mbox{if $f= e^{-1} $, } \\
0 & \mbox{otherwise. }
\end{array}
\right. 
\label{real}
\end{align}
The discrete-time QW with the Grover matrix ${\bf U}_{2m}$ as a time evolution matrix is the Grover walk with F-type on $G$. Then the $n \times n$ matrix ${\bf P}_{n} = [ P_{uv} ]_{u,v \in V(G)}$ is given by
\begin{align*}
P_{uv} =\left\{
\begin{array}{ll}
1/( \deg {}_G \ u)  & \mbox{if $(u,v) \in D(G)$, } \\
0 & \mbox{otherwise.}
\end{array}
\right.
\end{align*}
Note that the matrix ${\bf P}_n$ is the transition probability matrix of the simple symmetric RW on $G$. We introduce the {\em positive support} ${\bf F}^+ = [ F^+_{ij} ]$ of a real matrix ${\bf F} = [ F_{ij} ]$ as follows: 
\begin{align*}
F^+_{ij} =\left\{
\begin{array}{ll}
1 & \mbox{if $F_{ij} >0$, } \\
0 & \mbox{otherwise}.
\end{array}
\right.
\end{align*}
Ren et al. \cite{RenEtAl} showed that the Perron-Frobenius operator (or edge matrix) of a graph is the positive support $({}^{\rm{T}}{\bf U}_{2m})^+ $ of the transpose of its Grover matrix ${\bf U}_{2m}$, i.e., 
\begin{align}
{\bf Z} (G,u)^{-1} = \det \left( {\bf I}_{2m} -u( {}^{\rm{T}}{\bf U}_{2m})^+ \right)= \det \left( {\bf I}_{2m} -u {\bf U}_{2m} ^+ \right). 
\label{toruko01}
\end{align}
The Ihara zeta function of a graph $G$ is just a zeta function on the positive support of the Grover matrix of $G$. That is, the Ihara zeta function corresponds to the positive support version of the Grover walk (defined by the positive support of the Grover matrix ${\bf U}_{2m} ^+$) with F-type on $G$.

Now we propose another zeta function of a graph. Let $G$ be a simple connected graph with $m$ edges. Then we define a zeta function $ \overline{{\bf Z}} (G, u)$ of $G$ satisfying 
\begin{align}
\overline{{\bf Z}} (G, u)^{-1} = \det \left( {\bf I}_{2m} -u {\bf U}_{2m} \right).    
\label{toruko02}
\end{align}
In other words, this zeta function corresponds to the {\em Grover walk} (defined by the Grover matrix ${\bf U}_{2m}$) with F-type on $G$.

In this setting, Konno and Sato \cite{KonnoSato} presented the following result which is called the {\em Konno-Sato theorem}. 

\begin{theorem}
Let $G$ be a simple connected graph with $n$ vertices and $m$ edges. Then  
\begin{align}  
\overline{{\bf Z}} (G, u)^{-1} 
= \det ( {\bf I}_{2m} - u {\bf U}_{2m} )
=(1-u^2)^{m-n} \det \left( (1+u^2) {\bf I}_{n} -2u {\bf P}_n \right).
\label{wakatakakage1a}
\end{align}
\label{KS} 
\end{theorem}
It is noted that if we take $u = 1/\lambda$, then Eq. \eqref{wakatakakage1a} is rewritten as 
\begin{align*}
\det \left( \lambda {\bf I}_{2m} - {\bf U}_{2m} \right)
= ( \lambda {}^2 -1)^{m-n} \det \left( ( \lambda {}^2 +1) {\bf I}_n -2 \lambda {\bf P}_n \right).
\end{align*}

\section{Relation between QW and Absolute Zeta Function \label{sec04}}
Inspired by Eqs. \eqref{toruko01} and \eqref{toruko02}, we consider a zeta function $\zeta_{{\bf A}} (u)$ for an $N \times N$ matrix ${\bf A}$ defined by
\begin{align}  
\zeta_{{\bf A}} (u) = \left\{ \det \left( {\bf I}_{N} - u {\bf A} \right) \right\}^{-1}.
\label{mushiastui01}
\end{align}
If ${\bf A}$ is a regular matrix with its eigenvalues $\{\lambda_1, \ldots, \lambda_N\}$, then we easily see
\begin{align*}  
\zeta_{{\bf A}} \left( \frac{1}{u} \right)^{-1} 
&= \det \left( {\bf I}_{N} - \frac{1}{u} {\bf A} \right)
= \prod_{k=1}^N \left( 1 - \frac{\lambda_k}{u} \right)
= \left(\frac{1}{u} \right)^N \prod_{k=1}^N \left( u - \lambda_k \right)
\\
&
= \left(\frac{1}{u} \right)^N \left( \prod_{k=1}^N \lambda_k \right) \ (-1)^N \ \prod_{k=1}^N \left( 1 - \frac{u}{\lambda_k} \right)
\\
&
= (-u)^{-N} \ \det {\bf A} \ \left\{ \zeta_{{\bf A}^{-1}} \left( u \right) \right\}^{-1}. 
\end{align*}
Therefore we have the following result.
\begin{align}
\zeta_{{\bf A}} \left( \frac{1}{u} \right) = (-u)^{N} \ \left(\det {\bf A} \right)^{-1} \ \zeta_{{\bf A}^{-1}} \left( u \right).
\label{kakumei01}
\end{align}
On the other hand, we defined our zeta function $\overline{{\bf Z}} (G, u)$ based on a QW on the graph $G$ by
\begin{align}  
\overline{{\bf Z}} (G, u) = \left\{ \det \left( {\bf I}_{2m} - u {\bf U}_{2m} \right) \right\}^{-1},
\label{kakumei02}
\end{align}
where $G$ is a simple connected graph with $n$ vertices and $m$ edges (see Eq. \eqref{toruko02}). In order to clarify the dependence on a graph, from now on, we define ``${\bf U}_{2m}$ and $\overline{{\bf Z}} (G, u)$" by ``${\bf U}_{G}$ and $\zeta_{{\bf U}_{G}}$", respectively. So Eq. \eqref{kakumei02} is rewritten as 
\begin{align}  
\zeta_{{\bf U}_{G}} (u) = \left\{ \det \left( {\bf I}_{2m} - u {\bf U}_{G} \right) \right\}^{-1}.
\label{kakumei03}
\end{align}
Then it follows from Eqs. \eqref{mushiastui01}, \eqref{kakumei01} and \eqref{kakumei03} that 
\begin{align}
\zeta_{{\bf U}_{G}} \left( \frac{1}{u} \right) = (-u)^{2m} \ \left(\det {\bf U}_{G} \right)^{-1} \ \zeta_{{\bf U}_{G}^{-1}} \left( u \right).
\label{kakumei04}
\end{align}
We see that the definition of ${\bf U}_{G} (={\bf U}_{2m})$ given by Eq. \eqref{real} implies that each component of ${\bf U}_{G}$ is a real number. So ${\bf U}_{G}$ becomes an orthogonal matrix. Thus we have
\begin{align} 
{\bf U}_{G} ^{-1} = {}^{\rm{T}}{\bf U}_{G},
\label{kakumei05}
\end{align}
where $\rm{T}$ is the transposed operator. From Eq. \eqref{kakumei05}, we find
\begin{align*}  
\zeta_{{\bf U}_{G}} (u) 
&= \left\{ \det \left( {\bf I}_{2m} - u {\bf U}_{G} \right) \right\}^{-1} = \left\{ \det \left( {\bf I}_{2m} - u {}^{\rm{T}}{\bf U}_{G} \right) \right\}^{-1} 
\\
&= \left\{ \det \left( {\bf I}_{2m} - u {\bf U}_{G}^{-1} \right) \right\}^{-1} = \zeta_{{\bf U}_{G} ^{-1}} (u). 
\end{align*}
So we have
\begin{align}  
\zeta_{{\bf U}_{G}} (u) = \zeta_{{\bf U}_{G} ^{-1}} (u).
\label{kakumei06}
\end{align}
Moreover we get
\begin{align} 
\det {\bf U}_{G} \in \{-1,1\},
\label{kakumei07}
\end{align}
since ${\bf U}_{G}$ is an orthogonal matrix. Combining Eq. \eqref{kakumei04} with Eqs. \eqref{kakumei06} and \eqref{kakumei07}, we obtain the following result for our zeta function $\zeta_{{\bf U}_{G}} (u)$.
\begin{theorem}
Let $G$ be a simple connected graph with $n$ vertices and $m$ edges. Let ${\bf U}_{G} (={\bf U}_{2m})$ be a time evolution matrix of the Grover walk on $G$. Then we have
\begin{align}
\zeta_{{\bf U}_{G}} \left( \frac{1}{u} \right) = \det {\bf U}_{G} \ u^{2m} \ \zeta_{{\bf U}_{G}} \left( u \right),
\label{kakumei08}
\end{align}
where $\det {\bf U}_{G} \in \{-1,1\}$. 
\label{kakumei09}
\end{theorem}
Recall that $f$ is an absolute automorphic form of weight $D$ if $f$ satisfies
\begin{align*}
f \left( \frac{1}{u} \right) = C \ u^{-D} \ f(u)
\end{align*}
with $C \in \{ -1, 1 \}$ and $D \in \mathbb{Z}$. Therefore, from Theorem \ref{kakumei09}, we have an important property of our zeta function $\zeta_{{\bf U}_{G}} (u)$, that is, ``$\zeta_{{\bf U}_{G}} (u)$ is an absolute automorphic form of weight $- 2m$". Then $\zeta_{\zeta_{{\bf U}_{G}}} (s)$ is an absolute zeta function for our zeta function $\zeta_{{\bf U}_{G}} (u)$. In other words, we can consider ``the zeta function of a zeta function".

\section{Example \label{sec05}}
This section gives an example for $G=C_n$ (cycle graph) with $n$ vertices and $m=n$ edges. We defined ${\bf U}_{G}$ by ${\bf U}_{2m}$. Similarly, we put ${\bf P}_{G} = {\bf P}_{n}$. In this case, $2n \times 2n$ matrix ${\bf U}_{C_n} = {\bf U}_{2n}$ and $n \times n$ matrix ${\bf P}_{C_n} = {\bf P}_{n}$ are expressed as follows (see \cite{K1, K2, K5, Konno2022}, for instance):
\begin{align*}
{\bf U}_{C_n} 
&= {\bf U}_{2n} =
\begin{bmatrix}
O & P & O & \dots & \dots & O & Q \\
Q & O & P & O & \dots & \dots & O \\
O & Q & O & P & O & \dots & O \\
\vdots & \ddots & \ddots & \ddots & \ddots & \ddots & \vdots \\
O & \dots & O & Q & O & P & O \\
O & \dots & \dots & O & Q & O & P \\
P & O & \dots & \dots & O & Q & O
\end{bmatrix} 
, 
\\
{\bf P}_{C_n}
&={\bf P}_n = \frac{1}{2}
\begin{bmatrix}
0 & 1 & 0 & \dots & \dots & 0 & 1 \\
1 & 0 & 1 & 0 & \dots & \dots & 0 \\
0 & 1 & 0 & 1 & 0 & \dots & 0 \\
\vdots & \ddots & \ddots & \ddots & \ddots & \ddots & \vdots \\
0 & \dots & 0 & 1 & 0 & 1 & 0 \\
0 & \dots & \dots & 0 & 1 & 0 & 1 \\
1 & 0 & \dots & \dots & 0 & 1 & 0
\end{bmatrix} 
,
\end{align*}
where
\begin{align*}
P=
\begin{bmatrix}
1 & 0 \\
0 & 0 
\end{bmatrix} 
,
\quad
Q=
\begin{bmatrix}
0 & 0 \\
0 & 1 
\end{bmatrix} 
,
\quad
O=
\begin{bmatrix}
0 & 0 \\
0 & 0 
\end{bmatrix} 
.
\end{align*}
Let $\xi_k = 2 \pi k/n$ for $k=1, \ldots , n$. Then we have
\begin{align*}
{\rm Spec} \left( {\bf P}_{C_n} \right) 
={\rm Spec} \left( {\bf P}_n \right) 
= \left\{ [\cos \xi_k ]^{1} \ : \ k=1, \ldots , n \right\},
\end{align*}
where ${\rm Spec} ({\bf B})$ is the set of eigenvalues of a square matrix ${\bf B}$. More precisely, we also use the following notation: 
\begin{align*}
{\rm Spec} ({\bf B}) = \left\{ \left[ \lambda_1 \right]^{l_1}, \ \left[ \lambda_2 \right]^{l_2}, \ \ldots \ , \left[ \lambda_k \right]^{l_k} \right\},
\end{align*}
where $\lambda_j$ is the eigenvalue of ${\bf B}$ and $l_j \in \ZM_{>}$ is the multiplicity of $\lambda_j$ for $j=1,2, \ldots, k$. Note that if we apply the Konno-Sato Theorem (Theorem \ref{KS}) to ``${\rm Spec} \left( {\bf P}_{C_n} \right)$", then we obtain ``${\rm Spec} \left( {\bf U}_{C_n} \right)$" in the following way. 
\begin{align*}
{\rm Spec} \left( {\bf U}_{C_n} \right) 
= {\rm Spec} \left( {\bf U}_{2n} \right)
= \left\{ [e^{i \xi_k} ]^{1}, \  [e^{- i \xi_k} ]^{1} \ : \ k=1, \ldots , n \right\}. 
\end{align*}
For example, in the case of $n=4$ case, we easily see
\begin{align*}
{\rm Spec} \left( {\bf P}_{C_4} \right) 
&={\rm Spec} \left( {\bf P}_4 \right) 
= \left\{ \left[ \cos \left( 2 \cdot \pi/4 \right) \right]^{1}, \ \left[ \cos \left( 4 \cdot \pi/4 \right) \right]^{1}, \ \left[ \cos \left( 6 \cdot \pi/4 \right) \right]^{1} \ \left[ \cos \left( 8 \cdot \pi/4 \right) \right]^{1} \right\} \\
&
= \left\{ [-1]^1, \ [0]^2, \ [1]^1 \right\}, 
\\
{\rm Spec} \left( {\bf U}_{C_4} \right)
&={\rm Spec} \left( {\bf U}_8 \right)
= \left\{ \left[ e^{i (2 \cdot \pi/4) } \right]^{2}, \ \left[ e^{i (4 \cdot \pi/4) } \right]^{2}, \ \left[ e^{i (6 \cdot \pi/4) } \right]^{2}, \ \left[ e^{i (8 \cdot \pi/4) } \right]^{2} \right\}
\\
&
= \left\{ [-1]^2, \ \left[ i \right]^2, \ \left[ -i \right]^2, \ \left[ 1 \right]^2 \right\}.
\end{align*}
In this setting, we find 
\begin{align*}
\zeta_{{\bf U}_{C_n}} \left( u \right) ^{-1} 
&= \overline{{\bf Z}} (C_n, u)^{-1} = \det ( {\bf I}_{2n} - u {\bf U}_{C_n} )
\\
&=(1-u^2)^{n-n} \det \left\{ (1+u^2) {\bf I}_{n} -2u {\bf P}_{C_n} \right\}
\\
&= \prod_{k=1}^{n} \left\{ (1+u^2) -2u \cos \xi_k \right\}
= \prod_{k=1}^{n} \left(1 - e^{i \xi_k} u \right) \left(1 - e^{-i \xi_k} u \right) \\
&= \left\{ \prod_{k=1}^{n} \left(1 - e^{i \xi_k} u \right) \right\}^2 = \left( 1-u^n \right)^2.
\end{align*}
The third equality comes from the Konno-Sato Theorem (Theorem \ref{KS}). Thus we have
\begin{align}
\zeta_{{\bf U}_{C_n}} \left( u \right) = \frac{1}{\left( u^n - 1 \right)^2}.
\label{kirishima50}
\end{align}
Then we see that Eq. \eqref{kirishima50} is equivalent to Eq. \eqref{kirishima02}. Therefore it follows from  Eqs. \eqref{kirishima05} and \eqref{kirishima06} that $Z_{\zeta_{{\bf U}_{C_n}}} (w,s)$ and $\zeta_{\zeta_{{\bf U}_{C_n}}} (s)$ can be expressed as the multiple Hurwitz zeta function of order 2, $\zeta_2 (s, x, (\omega_1, \omega_2))$, and the multiple gamma function of order 2, $\Gamma_2 (x, (\omega_1, \omega_2))$, in the following way:
\begin{prop}
\begin{align}
Z_{\zeta_{{\bf U}_{C_n}}} (w,s) &= \zeta_2 (w, s+2n, (n,n)), 
\label{hokuseihou01}
\\
\zeta_{\zeta_{{\bf U}_{C_n}}} (s) &= \Gamma_2 (s+2n, (n,n)). 
\label{hokuseihou02}
\end{align}
\label{hokuseihou03}
\end{prop}

\par
Another derivation of Eqs. \eqref{hokuseihou01} and \eqref{hokuseihou02} is given by the following result based on Theorem 4.2 and its proof in Korokawa \cite{Kurokawa} (see also Theorem 1 in Kurokawa and Tanaka \cite{KT3}):
\begin{theorem}
For $\ell \in \mathbb{Z}, \ m(i) \in \mathbb{Z}_{>} \ (i=1, \ldots, a), \ n(j) \in \mathbb{Z}_{>} \ (j=1, \ldots, b)$, let 
\begin{align*}
f(x) = x^{\ell/2} \ \frac{\left( x^{m(1)} - 1 \right) \cdots \left( x^{m(a)} - 1 \right)}{\left( x^{n(1)} - 1 \right) \cdots \left( x^{n(b)} - 1 \right)}.
\end{align*}
Moreover we put 
\begin{align*}
\bec{n} = \left( n(1), \ldots, n(b) \right), \quad |\bec{n}| = \sum_{j=1}^b n(j), \quad m \left( I \right) = \sum_{i \in I} m(i), \quad |I| = \sum_{i \in I} 1.  
\end{align*}
Then we have 
\begin{align}
Z_f (w, s) 
&= \sum_{I \subset \{1, \ldots, a \}} (-1)^{a - |I|} \ \zeta_b \left( w, s - \frac{\ell}{2} + |\bec{n}| - m \left( I \right), \bec{n} \right),
\label{mkusatsu01}
\\
\zeta_f (s) 
&= \prod_{I \subset \{1, \ldots, a \}} \Gamma_b \left( s - \frac{\ell}{2} + |\bec{n}| - m \left( I \right), \bec{n} \right)^{ (-1)^{a - |I|}}.
\label{mkusatsu02}
\end{align}
\end{theorem}

\par
From Eq. \eqref{kirishima50}, our case is 
\begin{align*}
\zeta_{{\bf U}_{C_n}} \left( x \right) = f(x) = \frac{1}{\left( x^n - 1 \right)^2} =\frac{ x^n - 1}{\left( x^n - 1 \right)^3}.
\end{align*}
Noting that $\ell =0, \ a=1, \ m(1)=n, \ b=3, \ n(1)=n(2)=n(3)=n, \ \bec{n} =(n,n,n), \  |\bec{n}|=3n,$ it follows from Eq. \eqref{mkusatsu01} that
\begin{align*}
Z_{\zeta_{{\bf U}_{C_n}}} (w, s)
&= (-1)^{1-1}  \zeta_3 \left( w, s - \frac{0}{2} + 3n - n, (n,n,n) \right) 
\\
& \qquad \qquad + (-1)^{1-0}  \zeta_3 \left( w, s - \frac{0}{2} + 3n - 0, (n,n,n) \right) 
\\
&= \zeta_3 \left( w, s + 2n, (n,n,n) \right) - \zeta_3 \left( w, s + 3n, (n,n,n) \right)
\\
&= \zeta_2 \left( w, s + 2n, (n,n) \right).
\end{align*}
The last equality is obtained by a relation of the multiple Hurwitz zeta function (see proof of Theorem 3.5.1 in \cite{Kurokawa3}, for example). Similarly, by Eq. \eqref{mkusatsu02}, we see  
\begin{align*}
\zeta_{\zeta_{{\bf U}_{C_n}}} (s)
&= \Gamma_3 \left( s - \frac{0}{2} + 3n - n, (n,n,n) \right)^{(-1)^{1-1}}  
\\
& \qquad \qquad \times \Gamma_3 \left( s - \frac{0}{2} + 3n - 0, (n,n,n) \right)^{(-1)^{1-0}}   
\\
&= \Gamma_3 \left( s + 2n, (n,n,n) \right) \times \Gamma_3 \left( s + 3n, (n,n,n) \right)^{-1}
\\
&= \Gamma_2 \left( s + 2n, (n,n) \right).
\end{align*}
The last equality comes from a relation of the multiple gamma function (see Theorem 3.5.1 in \cite{Kurokawa3}, for example). Therefore, we have the same conclusion as Eqs. \eqref{hokuseihou01} and \eqref{hokuseihou02} in Proposition \ref{hokuseihou03}. We should remark that Theorem 4.1 in \cite{Kurokawa}  (see also Theorem 1 in \cite{KT3}) implies that $f$ is an absolute automorphic form of weight $D$:
\begin{align*}
f \left( \frac{1}{x} \right) = C x^{-D} f(x)
\end{align*}
with $C = (-1)^{a-b} \in \{-1, 1\}$ and $D = \ell + |\bec{m}| - |\bec{n}| \in \mathbb{Z}$, where 
\begin{align*}
\bec{m} = \left( m(1), \ldots, m(a) \right), \quad |\bec{m}| = \sum_{i=1}^a m(i).  
\end{align*}
In our case, it follows from $C=(-1)^{1-3}=1$ and $D = \ell + |\bec{m}| - |\bec{n}|= 0 + n - 3n = -2n$ that
\begin{align*}
f \left( \frac{1}{x} \right) = x^{2n} f(x).
\end{align*}
That is,
\begin{align}
\zeta_{{\bf U}_{C_n}} \left( \frac{1}{x} \right) = x^{2n} \zeta_{{\bf U}_{C_n}} (x).
\label{aserora01}
\end{align}
On the other hand, Eq. \eqref{kakumei08} in Theorem \ref{kakumei09} yields
\begin{align}
\zeta_{{\bf U}_{C_n}} \left( \frac{1}{x} \right) = \det {\bf U}_{C_n} \ x^{2m} \ \zeta_{{\bf U}_{C_n}} \left( x \right) = x^{2n} \ \zeta_{{\bf U}_{C_n}} \left( x \right),
\label{aserora02}
\end{align}
since $m=n$ and $\det {\bf U}_{C_n}=1$. So we confirm that both Eqs. \eqref{aserora01} and \eqref{aserora02} are same.

Furthermore, in a similar fashion of the derivation of Eq. \eqref{hokuseihou02} in Proposition \ref{hokuseihou03}, we have the following result called the {\em functional equation}:
\begin{align}
\zeta_{\zeta_{{\bf U}_{C_n}}} (-2n - s) = S_2 (s+2n, (n,n)) \ \zeta_{\zeta_{{\bf U}_{C_n}}} (s),
\label{kaibutsu02}
\end{align}
where the {\em multiple sine function of order $r$}, $S_r (x, (\omega_1, \ldots, \omega_r))$, is defined by 
\begin{align*}
S_r (x, (\omega_1, \ldots, \omega_r))
= \Gamma_r (x, (\omega_1, \ldots, \omega_r))^{-1} \ \Gamma_r (\omega_1+ \cdots + \omega_r - x, (\omega_1, \ldots, \omega_r))^{(-1)^r}.
\end{align*}
Here $\Gamma_r (x, (\omega_1, \ldots, \omega_r))$ is the the multiple gamma function of order $r$. As for the multiple sine function, see \cite{Kurokawa3, Kurokawa, KT3}, for example.

\section{Conclusion \label{sec06}}
In this paper, first we dealt with a zeta function $\zeta_{{\bf U}_{G}} (u)$ determined by ${\bf U}_{G}$ which is a time evolution matrix of the Grover walk on $G$, where $G$ is a simple connected graph with $n$ vertices and $m$ edges. Then we proved that $\zeta_{{\bf U}_{G}} (u)$ is an absolute automorphic form of weight $- 2m$ (Theorem \ref{kakumei09}). After that we considered the absolute zeta function $\zeta_{\zeta_{{\bf U}_{G}}} (s)$ for our zeta function $\zeta_{{\bf U}_{G}} (u)$. As an example, we computed $\zeta_{\zeta_{{\bf U}_{G}}} (s)$ for the cycle graph $G = C_n$ with $n$ vertices and $m=n$ edges, and showed that it is expressed as the multiple gamma function of order 2. Finally, we obtained the functional equation for $\zeta_{\zeta_{{\bf U}_{C_n}}} (s)$ using the multiple sine function of order 2. One of the interesting future problems might be to obtain $\zeta_{\zeta_{{\bf U}_{G}}} (s)$ for other graphs except for the cycle graph.





\begin{thebibliography}{99}




\bibitem{Andrews1999} 
Andrews, G. E., Askey, R., Roy, R.: 
Special Functions. 
Cambridge University Press (1999)


\bibitem{Bass}
Bass, H.: 
The Ihara-Selberg zeta function of a tree lattice. 
Internat. J. Math. {\bf 3}, 717-797 (1992)


\bibitem{CC} 
Connes, A., Consani, C.: 
Schemes over $\mathbb{F}_1$ and zeta functions.
Compositio Math. {\bf 146}, 1383--1415 (2010)


\bibitem{GZ}
Godsil, C., Zhan, H.:
Discrete Quantum Walks on Graphs and Digraphs.
Cambridge University Press (2023)


\bibitem{Ihara}
Ihara, Y.: 
On discrete subgroups of the two by two projective linear group 
over $p$-adic fields. 
J. Math. Soc. Japan {\bf 18}, 219--235 (1966)


\bibitem{K1}
Komatsu, T., Konno, N., Sato, I.: 
Grover/Zeta Correspondence based on the Konno-Sato theorem.
Quantum Inf. Process. {\bf 20}, 268 (2021) 


\bibitem{K2}
Komatsu, T., Konno, N., Sato, I.: 
CTM/Zeta Correspondence. 
Quantum Stud.: Math. Found. 
{\bf 9}, 165--173 (2022)


\bibitem{K5}
Komatsu, T., Konno, N., Sato, I.: 
Walk/Zeta Correspondence. 
J. Stat. Phys. {\bf 190}, 36 (2023) 


\bibitem{Konno2008}
Konno, N.: 
Quantum Walks. In: Quantum Potential Theory, Franz, U., and Schurmann,
M., Eds., Lecture Notes in Mathematics: Vol. 1954, pp.309--452, Springer-Verlag, Heidelberg (2008)


\bibitem{Konno2022}
Konno, N.: 
An analogue of the Riemann Hypothesis via quantum walks. 
Quantum Stud.: Math. Found. 
{\bf 9}, 367--379 (2022)


\bibitem{KonnoSato} 
Konno, N., Sato, I.: 
On the relation between quantum walks and zeta functions. 
Quantum Inf. Process. {\bf 11}, 341--349 (2012)


\bibitem{KF1}
Kurokawa, N.: 
Zeta functions over $\mathbb{F}_1$.
Proc. Japan Acad. Ser. A Math. Sci. {\bf 81}, 180--184 (2005)


\bibitem{Kurokawa3}  
Kurokawa, N.:
Modern Theory of Trigonometric Functions.
Iwanami Publication, Tokyo (2013) in Japanese.


\bibitem{Kurokawa}  
Kurokawa, N.:
Theory of Absolute Zeta Functions.
Iwanami Publication, Tokyo (2016) in Japanese.


\bibitem{KO}
Kurokawa, N., Ochiai, H.: 
Dualities for absolute zeta functions and multiple gamma functions.
Proc. Japan Acad. Ser. A Math. Sci. {\bf 89}, 75--79 (2013)


\bibitem{KT3}
Kurokawa, N., Tanaka, H.: 
Absolute zeta functions and the automorphy. 
Kodai Math. J. {\bf 40}, 584--614 (2017)


\bibitem{KT4}
Kurokawa, N., Tanaka, H.: 
Absolute zeta functions and absolute automorphic forms.
J. Geom. Phys. {\bf 126}, 168--180 (2018)




\bibitem{ManouchehriWang}
Manouchehri, K., Wang, J.: 
Physical Implementation of Quantum Walks.
Springer, New York (2014)


\bibitem{Norris}
Norris, J. R.: Markov Chains. 
Cambridge University Press, Cambridge (1997)


\bibitem{Portugal} 
Portugal, R.: 
Quantum Walks and Search Algorithms, 2nd edition. 
Springer, New York (2018)


\bibitem{RenEtAl}  
Ren, P., Aleksic, T., Emms, D., Wilson, R. C., Hancock, E. R.: 
Quantum walks, Ihara zeta functions and cospectrality in regular graphs.  
Quantum Inf. Process. {\bf 10}, 405--417 (2011)   


\bibitem{Soule}
Soul\'e, C.: 
Les vari\'et\'es sur le corps \`a un \'el\'ement.  
Mosc. Math. J. {\bf 4}, 217--244 (2004) 


\bibitem{Spitzer}  
Spitzer, F.: 
Principles of Random Walk, 2nd edition. 
Springer, New York (1976)


\bibitem{Venegas} 
Venegas-Andraca, S. E.: 
Quantum walks: a comprehensive review. 
Quantum Inf. Process. {\bf 11}, 1015--1106 (2012)




\end{thebibliography}
\end{document}